\documentstyle[preprint,prl,aps]{revtex}

\newcommand{\Kr}[1]{\left( #1\right)}

\newcommand{\ve}{\varepsilon}
\newcommand{\vf}{v_{\rm F}}
\newcommand{\kf}{k_{\rm F}}
\newcommand{\ab}{a_{\rm B}}

\begin{document}


\title{Spin and Charge Luttinger-Liquid Parameters of the
One-Dimensional Electron Gas}

\author{C.E. Creffield$^1$, Wolfgang H\"ausler$^{1,2}$, and
A.H. MacDonald$^3$}

\address{$^1$ Dept.\ of Physics, King's College London,
Strand, London, WC2R~2LS, U.K.\\
$^2$ I.~Institut f\"ur Theoretische Physik der Universit\"at
Hamburg, Jungiusstr.~9, D-20355 Hamburg, Germany\\
$^3$ Dept.\ of Physics, Indiana University, 701 E Third Street,
Swain Hall-West 117, Bloomington, IN 47405, U.S.A.}

\date{\today}

\maketitle

\begin{abstract}
Low-energy properties of the homogeneous electron
gas in one dimension are completely described by the group velocities of
its charge (plasmon) and spin collective excitations. Because of
the long range of the electron-electron interaction, the
plasmon velocity is dominated by an electrostatic contribution and
can be estimated accurately. In this Letter
we report on Quantum Monte Carlo simulations which demonstrate that
the spin velocity is substantially decreased by interactions
in semiconductor quantum wire realizations of the one-dimensional
electron liquid.

\end{abstract}
\vskip2pc
\noindent PACS~: 71.10.Pm, 71.15.Pd
\newpage
The homogeneous electron liquid (HEL), historically important
\cite{bloch,wignerseitz} as a simple model of a metal, continues
to be of interest both in three-dimensions (3D) \cite{ortiz} and
in systems of reduced dimensionality. The quantum ground state
of this model system is determined by a competition between
kinetic and interaction energies, and depends only on $r_s$, the
radius in atomic units of a sphere containing one electron.
When the density is high ($r_s \lesssim 1$), the ground state
energy and, more interestingly, the parameters which
characterize the system's low energy excitations, can be
evaluated using semi-analytic perturbative
techniques \cite{electrongastheory}. At lower densities (larger
$r_s$), correlations are strong and numerical Quantum Monte Carlo
(QMC) calculations are required \cite{qmc}. In this article we
report on QMC calculations for the 1D \cite{senatore} case.

Unlike their higher dimensional counterparts, 1D interacting fermion
systems are not Fermi liquids; instead they exhibit the
low-energy phenomenology common to all 1D fermion systems, often
referred to as Tomonaga-Luttinger (TL) liquids \cite{haldane,reviews}.
Any TL liquid is completely specified by four parameters: the
charge and spin collective excitation velocities $v_{\rho}$ and
$v_{\sigma}$, and the correlation exponents $K_{\rho}$ and $K_{\sigma}$.
In the absence of interactions $v_{\rho}$ and $v_{\sigma}$ reduce
to the Fermi velocity, $\vf$, and the correlation exponents are equal to
$1$. Many quantities of immediate physical
relevance can be expressed in terms of the TL-parameters
\cite{reviews,kanefisher}.

Symmetry and other considerations can reduce the number of
independent TL parameters. The Galilean invariance of continuum
models simplifies the charge sector, since the product
\cite{drude} $v_{\rho} K_{\rho} = \vf$ is not altered by
interactions. Spin-rotational invariance simplifies the
spin-sector since it requires $K_{\sigma}=1$ \cite{reviews}.
The low energy physics of the HEL is thus specified by
$v_{\rho}$ and $v_{\sigma}$.

In this Letter we use QMC simulations to estimate, for the first
time, $v_{\rho}$ and $v_{\sigma}$ \cite{othermodels}.
The values we find for $v_{\rho}$ in typical experimental
circumstances are consistent with the large enhancements
predicted perturbatively. Confirming this relation can be
useful for a microscopic understanding of the exponents deduced
from power laws in the transport properties at finite voltages
or temperatures \cite{bockrath}. In the spin sector most works
leave $v_{\sigma}$ unspecified or assume that $v_{\sigma}$ would
not deviate much from $\vf$. Taking the model originally
proposed by Luttinger \cite{luttinger} with left and right going
particles treated as distinguishable, the spin velocity indeed
stays unrenormalized, $v_{\sigma}=\vf$ \cite{overhauser}.
Reliable knowledge of $v_{\sigma}$ is important to understand
the operation of devices where spin instead of charge is
transported \cite{tsukagoshi99}, such as the `spin transistor'
\cite{spintransistor}. We find that $v_{\sigma}$ is not
described well by perturbative estimates, and, contrary to
common wisdom, is strongly reduced by interactions.

The 1D HEL\cite{wires,nanos} is realized in semiconductor
systems with carriers electrostatically or chemically confined along
a line, and with densities sufficiently low to quantize transverse motion,
{\it i.e.} to realize the single-channel limit of a
quantum wire.
In 1D, the singularity of the $1/r$ repulsive interaction
between electrons is cut off at short distances
by the transverse width $d$ of the lowest subband wavefunction.
Usually the interaction is also cut off at large distances $R$
due to screening by mobile external electrons in gates which help to
define the electron channel. We use the convenient
form for the electron-electron interaction:
\begin{equation}
\label{intera}
V(|x|)=e^2\Kr{\frac{1}{\sqrt{x^2+d^2}}-
\frac{1}{\sqrt{x^2+d^2+4R^2}}}\;.
\end{equation}
This form is suggested by an image charge model of
remote screening. We consider a system to be a 1D HEL
provided that only one subband is occupied and $R\gg r_s\ab$,
where $\ab$ is the host semiconductor Bohr radius.
In 1D $r_s$ is related to the density by
$n = 1 / (2 r_s \ab)$, and has to exceed a minimum value
$r_s^{min}$ to suppress higher subband occupations.
For a parabolic transverse confining potential
$r_s^{min} = (\pi / 4 \sqrt{2}) (d/\ab) \sim 0.55 (d/\ab)$.
Values of $d/\ab$ and $R/\ab$ for a particular sample can be
estimated from measured subband energy separations and the
sample layout respectively. For the numerical calculations
described below, we use $d = 0.5 \ab$ and $R = 7.07 \ab$. This
value of $d$ is probably somewhat smaller than that which can be
achieved at present in samples where interactions dominate disorder,
while the value of $R$ is somewhat smaller than a typical one,
since we want to study the crossover to short-range interaction
physics which occurs when $r_s$ exceeds $R/\ab$.

For the QMC simulations a lattice representation
\begin{eqnarray}
H &=& -\frac{\hbar^2}{m^{\ast} L^2} \ \frac{M^2}{2}
\Bigl[\sum_{\sigma \langle i,j \rangle}
(c^{\dagger}_{i \sigma} c^{}_{j \sigma} + \mbox{h.c.})- \ 2 N \Bigr]
\nonumber \\
+&&\sum_i V_{i i} \ n_{i \uparrow} n_{i \downarrow}
+ \ \sum_{i < j} V_{i j} \ n_i n_j
\label{model}
\end{eqnarray}
is particularly convenient since the standard `world-line' QMC
algorithm \cite{hirsch} can be employed, and correlation
functions diagonal in number operators, such as the on-site
charge ($n_{i \uparrow} + n_{i \downarrow}$), are simple to
calculate. The number of lattice points used in our simulations
ranged from $M=32$ to $M=64$ on a chain of length $L$ (with
periodic boundary conditions), while 960 time-slices divided the
Trotter-axis. The inverse temperature, $\beta$, was set to a
high value, $\beta=48 \ 2m^*L^2/M^2$, to ensure that the
simulation sampled the ground state properties of the model.
The continuum limit is recovered at particle numbers $N$ such
that $N/M\ll 1$, for which the quadratic single-particle
spectrum is regained. In Eq.(~\ref{model}) $V_{ij}=V(L|i-j|/M)$.

As a test of our simulation procedure and our model, we have calculated the
density dependence of the ground state energy. The results presented
below are based on 4000 samples, drawn from $\sim 10^5$
configurations, and are plotted as a function of
$\kf d = \pi(d / \ab)/4 r_s$, where $\kf$ is the Fermi momentum.
To rule out ergodic `sticking' we compared
results from different starting configurations, and found that
in all cases the energies converged to the same values within
statistical errors. A careful finite-size scaling analysis for the
ground state energy density $E_0$ was carried out based on the form
\cite{cardy} $E_0(L) = E_0(L=\infty)-c/L^2$, which fitted the
simulation results extremely well.

Since a full finite-size analysis for every value of $\kf d$ would
be too time consuming, we chose instead to estimate the
corrections by interpolating linearly between the finite-size
corrections obtained at $\kf d=0.079$ $(r_s =4.97)$ and
$\kf d=0.5$ ($r_s =0.78$).
Additional finite-size analyses at $\kf d=0.132$ and
$\kf d=0.314$ confirmed the accuracy of this interpolation.

\begin{figure}
\vspace*{7.5cm}

\includegraphics{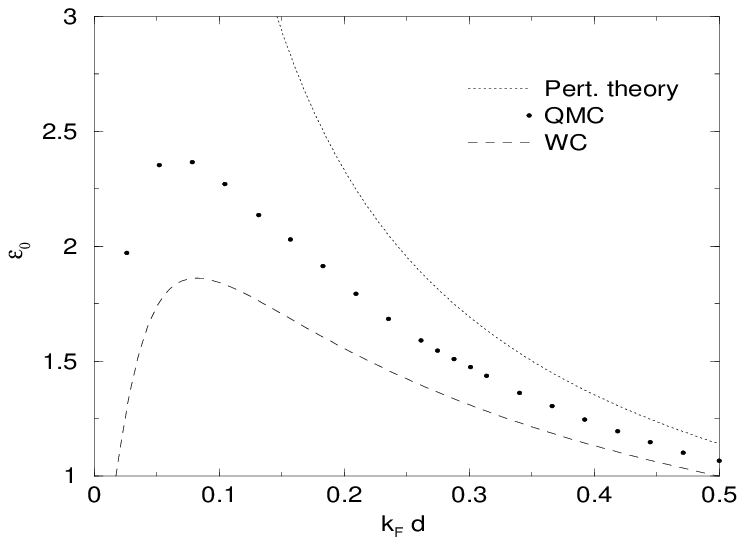}\caption[]{\label{e0}
QMC dimensionless ground state energy density,
($\ve_0=E_0/\vf\kf^2$) of quantum wires versus $\kf d$
for $R/d=14.14$ and $d/\ab=0.5$. Also included are~:
the Hartree-Fock (Eq.~(\ref{pert})) and
harmonic (dashed, cf.\ Eq.~(\ref{clfl})) approximations
to the energy.}
\end{figure}

Fig.~\ref{e0} shows the dimensionless ground state energy density
$\ve_0(\kf)=E_0m^*/\kf^3$, which has the value
$1/3\pi \sim 0.106 $ in the absence of interactions.
For comparison we have also plotted the first order perturbation
(Hartree-Fock) theory result:
\begin{eqnarray}\label{pert}
\ve_0^{\rm pert}&=&\frac{1}{3\pi}+\frac{2}{\pi^2 \hbar\vf}\hat{V}(0)
\\
&-&\frac{1}{2\pi^2\hbar\vf\kf^2}\int_0^{2\kf}{\rm d}k\;(2\kf-k)\hat{V}(k)
\nonumber
\end{eqnarray}
where the three terms on the right hand side are kinetic energy, mean-field
electrostatic energy, and exchange energy contributions, and
$\hat{V}(k)$ is the Fourier transform of the interaction
(\ref{intera}).

We also include in Fig.~\ref{e0} the harmonic lattice
approximation to the energy of an electron crystal state:
\begin{equation}\label{clfl}
E_0^{\rm wc}=E_0^{\rm cl}+\frac{1}{2}\int_{-\kf}^{\kf}
\frac{{\rm d}k}{2\pi}\;\omega(k)
\end{equation}
where $E_0^{\rm cl}=\frac{1}{2L}\sum_{i\ne j}V(|i-j|/n)$
is the classical energy,
$\omega^2(k)=\frac{1}{m^*}\sum_{j=1}^{\infty}V''(j/n)(1-\cos
jk/n)$ is the dispersion of harmonic
fluctuations, and primes denote derivatives w.r.t.\ the arguments.
This latter approximation sets a lower bound to the true ground
state energy since the quartic term of the Coulomb interaction
is positive when expanded as a power series.
Conversely, the variational nature of the Hartree-Fock estimate
guarantees that it is an upper bound. We see in Fig.~\ref{e0} that
in the high-density regime, where the electrostatic term dominates the energy,
these bounds limit the gound state energy to a narrow interval. At smaller
densities, when $r_s\gtrsim 2$, we cross over to the
short-range interaction regime within which
Hartree-Fock estimate fails especially badly.

\begin{figure}
\vspace*{5.5cm}

\includegraphics{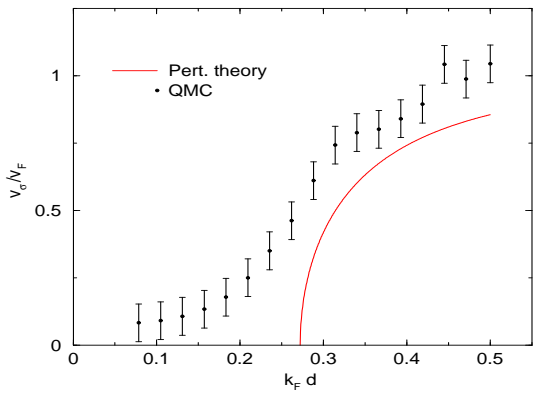}\caption[]{\label{vsigma}
Spin velocity, normalized to the Fermi velocity. The perturbative
result is given by the solid line.
}
\end{figure}

The most striking result of this work, the spin-velocity TL parameter,
is presented in Fig.~\ref{vsigma}. This quantity was calculated
using the TL-theory
expression\cite{reviews} for the static homogeneous spin susceptibility
$\chi(q\to 0,\omega = 0) = (2/\pi) K_{\sigma}/v_{\sigma}$,
with $K_{\sigma}=1$. The static spin-susceptibility is readily extracted
from the simulations by integrating
the QMC Matsubara spin-density -- spin-density
correlation function over imaginary time. In the same
way, the compressibility $\kappa=(2/\pi)(K_{\rho}^2/\vf)$ can be derived
from the integral over imaginary time of the Matsubara
density-density correlation function. The values calculated in
this manner agree with those obtained from the second derivative of
the the ground state energies (see below),
confirming the reliability of this procedure. The spin
correlation function, however, was found to be considerably
noisier than its density counterpart, particularly at lower density where
correlations are stronger. As a result the spin data are
subject to larger statistical errors. The QMC estimate can be
compared with the perturbative generalized-random-phase
approximation\cite{grpa} estimate,
$v_{\sigma}^{\rm GRPA}/\vf= (1-\hat{V}(2\kf)/\pi\vf)^{1/2}$,
also shown in Fig.~\ref{vsigma}. The GRPA spin-velocity goes to zero as the
ferromagnetic instability predicted by the Hartree-Fock energy approximation
is approached. (This occurs at $r_s \sim 1.44$ for $d = 0.5 \ab$.)
In 2D\cite{stern} the instability occurs at
$r_s = \pi/\sqrt{2} \sim 2.22$, and in 3D\cite{bloch} at
$r_s = (9 \pi^4/4)^{1/3} \sim 6.03$.

Ferromagnetic instabilities predicted by the Hartree-Fock approximation
have provided
an important motivation for the Monte Carlo calculations in higher
dimensions, where it has been established\cite{ortiz,qmc} that the
transition does not occur until substantially larger $r_s$ values are
reached, if at all. In 1D the Lieb-Mattis theorem\cite{mattis62} guarantees
that the ferromagnetic transition does not occur
at any $r_s$, in disagreement with conclusions based on recent
density-functional calculations\cite{gold}, which give results similar to
the Hartree-Fock approximation results quoted here. Nevertheless, this
instability may be taken as a marker for the crossover from weak to
strong correlations which, we note, occurs at substantially smaller values
of $r_s$ in reduced dimension systems.

For the quantum wire system\cite{wires} of Yacoby {\em et al.},
we estimate $d/\ab \sim 0.7$ and take the `typical' density to be one fifth
of that at which the second subband is occupied. This corresponds to
$r_s \sim 1.93 $ and $\kf d \sim 0.28$, close to the value at which
ferromagnetism is predicted in the Hartree-Fock approximation.
Near this density, our QMC calculations show a paramagnetic state,
albeit one with a spin velocity reduced by a factor of more than two
compared to the Fermi velocity.
Already at this density, a fractional spin-polarization $\xi$ can be produced
by laboratory fields: $\xi =
(S(\xi) r_s^2/2) (g \mu_B B)/(\hbar^2/ 2 m^* \ab^2)$ where $S(\xi)$ is
the enchancement due to interactions. (For $\xi \to 0, \quad S(\xi) =
v_F/v_{\sigma}$.) In GaAs $(g \mu_B B)/(\hbar^2/ 2 m^* \ab^2) \sim
0.0052 B$ [Tesla], so that for $r_s \sim 1.9$,
$\xi \sim 0.2$ at $B = 20$ Tesla, even when interaction
enhancement is neglected. Note that $S(\xi)$ grows rapidly
with further density decreases. We conclude that reduced magnetic
stiffness, due to correlations which keep electrons apart, will lead to
substantial spin-polarization in quantum wires at routinely available
magnetic fields, and must play a role in quantum transport experiments,
like those of Thomas et al.\cite{thomas}. More directly, spin
velocities can be measured using inelastic light scattering in
depolarized configuration \cite{raman} or, with
ferromagnetic contacts attached \cite{tsukagoshi99}, using time
resolved techniques \cite{haug}.

In closing we discuss the charge physics of semiconductor quantum wires.
The thermodynamic relationship between the compressibility
$\kappa$ and the LL-parameter,
\begin{eqnarray}\label{ke0}
K_{\rho}^{-2}&=&(2/\pi\vf)\kappa^{-1}
\\
&=&\frac{\pi}{2}[\kf^2\ve_0''(\kf)+
6(\kf\ve_0'(\kf)+\ve_0(\kf))]
\nonumber
\end{eqnarray}
enables very accurate estimates to be made for $v_{\rho}$. A
similar relation has been used previously for Hubbard, and
Sutherland models \cite{othermodels}. Eq.~(\ref{ke0}) is
particularly suited to the QMC approach since thermodynamic
quantities converge more rapidly than dynamical ones, and can
also be corrected for finite-size effects more easily. The QMC
energies were smoothed by a cubic spline fit to avoid amplifying
any small irregularities in the stochastic data in performing
the numerical differentiations in (\ref{ke0}).

\begin{figure}
\vspace*{7.5cm}

\includegraphics{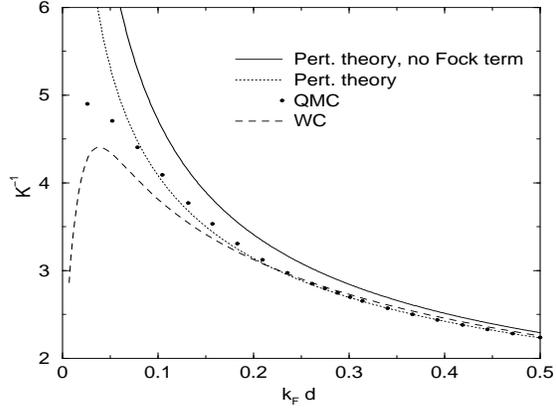}\caption[]{\label{kr}
$K_{\rho}^{-1}$ versus $\kf d$. The same approximations are
included as in Fig.~\ref{e0} and in addition the result without
the negative exchange term in (\ref{pertk}).}
\end{figure}

In Fig.~\ref{kr} the QMC correlation exponents are compared
with GRPA and harmonic lattice values. The GRPA compressibility
can be obtained from the corresponding Hartree-Fock energy expression
which leads to
\begin{equation}
\label{pertk}
K_{\rho}^{\rm GRPA}=[1+(2\hat{V}(k=0)-\hat{V}(k=2\kf))/\pi\vf]^{-{1\over2}}.
\end{equation}
We would like to emphasize that the good agreement of all approximations
with the exact result seen in Fig.~\ref{kr} down to quite low
densities is not expected a priori. None of the approximations
provides a rigorous bound to $K_{\rho}$. In the language of
the renormalization group approach to 1D interacting fermion systems,
irrelevant operators, for example
backscattering in the spin sector, can in principle alter the
value of low energy TL-parameters. Apparently the renromalization
corrections to $K_{\rho}$,
which has not yet been studied quantitatively for the HEL,
is quite small at moderate and high densities.
At low densities the GRPA estimate for $K_{\rho}^{-1}$ is
proportional to $\vf^{-1} \propto r_s$. The approximation therefore fails to
reproduce either the maximum seen in the harmonic lattice approximation,
or the precursor of this maximum visible in the plotted QMC data.
It has been conjectured that $1/K_{\rho}\to 2$ as $\kf\to 0$ for
any finite-range interaction\cite{mfa}.
At very small densities, $\kf<d/R^2$, the harmonic approximation
$1/K_{\rho}^{\rm wc}\sim\kf^{1/4}$ also fails, underestimating the residual
kinetic energy and violating the condition that $K_{\rho}<1$ for
repulsive interactions.
A closer inspection of the QMC data in Fig.~\ref{kr} reveals
that $1/K_{\rho}$ actually {\em exceeds} the perturbative estimate
before it levels off towards the maximum. In this regime the
system is clearly stiffer (\ref{ke0}) than perturbation
theory suggests. This behavior is also seen in higher dimensions, where
the inverse compressibility is underestimated by the GRPA.
A quantum wire model with a smaller $1/R$ would extend this
regime and make the enhancement even more pronounced.

In conclusion, we have established by means of QMC the regime of validity of
the widely used perturbative expression for the charge correlation exponent
$K_{\rho}$ and have found that spin velocities in typical experimental
realizations are significantly reduced by interactions. This finding
is of direct relevance to the upcoming `spintronic' devices using the HEL.

We would like to thank Sarben Sarkar and particularly John Jefferson for
valuable encouragement and Gaitano Senatore for helpful discussions.
CEC is grateful to the Leverhulme Foundation and WH to the
British EPSRC for financial support. The work of AHM was supported
by the National Science Foundation under grant
DMR-9714055.

\begin{raggedright}
\end{raggedright}
\vspace{3cm}

\newpage


\begin{thebibliography}{10}
\bibitem{drude} This property is one consequence of the invariance
of the interaction energy under a shift of the
many-electron state in momentum space, producing a Drude weight in
the conductivity which is independent of interactions.
\bibitem{grpa} The generalized random-phase-approximation for
the spin-susceptibility is obtained by including ladder diagram
vertex corrections to the single-loop susceptibility calculated
with Hartree-Fock Greens functions.
\end{thebibliography}
\end{document}